\documentclass[a4paper,11pt]{article}
\usepackage{pos}

\newcommand{\eq}{\begin{equation}}
\newcommand{\en}{\end{equation}}
\newcommand{\Tr}{\ensuremath{\mathrm{Tr}}}
\newcommand{\SU}{\mathrm{SU}}

\title{Effective string description of the reconfined phase in the trace deformed $\SU(2)$ Yang-Mills theory in (2+1) dimensions.}
\ShortTitle{EST description of  reconfinement.}

\author[a]{Claudio Bonati}
\author*[b]{Michele Caselle}
\author[c]{Alessio Negro}
\author[b]{Dario Panfalone}
\author[b]{Lorenzo Verzichelli}

\affiliation[a]{Department of Physics, University of Pisa and INFN, Pisa,\\
 Largo Pontecorvo 3, I-56127 Pisa, Italy.}

\affiliation[b]{Department of Physics, University of Turin and INFN, Turin,\\
  Via Pietro Giuria 1, I-10125 Turin, Italy.}

\affiliation[c]{Helmholtz-Institut f\"ur Strahlen- und Kernphysik, University of Bonn, \\
    Nussallee 14-16, 53115 Bonn, Germany and \\
    Bethe Center for Theoretical Physics, University of Bonn,\\
    Nussallee 12, 53115 Bonn, Germany}

\emailAdd{michele.caselle@unito.it}

\abstract{We study the behaviour of the flux tube in the reconfined phase of the trace deformed $\SU(2)$ Yang-Mills theory in $(2+1)$ dimensions. In this phase the Polyakov loop has a vanishing expectation value (and center symmetry is recovered) even at high temperatures. We study, by means of numerical simulations, the confining potential between two Polyakov loops. We show that its behaviour is very different from that of usual confining gauge models and shows a remarkable agreement with the predictions of the so called "rigid string" in the limit in which the rigidity term (\textit{i.e.} a term proportional to the square of the extrinsic curvature of the string) is very large and is the dominant contribution in the action.}

\FullConference{The 41st International Symposium on Lattice Field Theory (LATTICE2024)\\
 28 July - 3 August 2024\\
Liverpool, UK\\}

\begin{document}
\maketitle

\section{Introduction}
Understanding confinement is one of the main open problems in Yang-Mills (YM) theories. 
To address this problem it was proposed long ago  to  compactify the theory on a ${\bf R}^3 \times S$ manifold. 
The compactification radius sets the energy scale of the theory and the hope is, by squeezing the radius, to reach a weak coupling regime where perturbative methods can be used. 
As it is well known this approach, in its simplest version, is too naive because at some critical value $N_{t,c}$ of the compactification radius $N_t$ the model undergoes a deconfinement transition.
During the past fifty years several attempts have been made to avoid this deconfining transition but none of them led to a reliable solution of the problem. Recently this line of resarch attracted  new interest thanks to a new original proposal. The idea is to modify the YM action by adding a {\sl "trace deformation"} term so as to keep the expectation value of the Polyakov loop to zero and thus the model in the confined phase even for temperatures above the deconfinement transition \cite{Unsal:2008ch,Myers:2007vc}.

The question is thus if this {\sl "reconfined"} phase of the model shares the same properties with the original confining phase.
Some properties seem to be conserved: the glueball spectrum \cite{Athenodorou:2020clr}, the localization/delocalization transition of the Dirac eigenvalues \cite{Bonati:2020lal} and the $\theta$ dependence of the free energy \cite{Bonati:2018rfg}.
Also the properties related to the condensation of monopoles are substantially equivalent in the origianl YM and the trace deformed theory \cite{Bonati:2020lal}.

It seems clear that in order to answer this question in an unambiguous way one should directly study the behaviour of the confining flux tube in the reconfined phase and compare it with that of the ordinary confining phase. 
This is precisely our goal. We address the problem in the particular case of the the $\SU(2)$ pure gauge model in (2+1) dimensions since it is the simplest LGT with a non-abelian continuous gauge group and is thus a perfect laboratory to
test large distance, non perturbative, features of YM theories with a relatively small numerical effort.
Another advantage of this choice is that this model has been the subject of several studies in the past \cite{Ambjorn:1984me, Teper:1998te, Caselle:2004er, Caselle:2011vk, Bringoltz:2006zg, Brandt:2010bw, Brandt:2017yzw, Brandt:2018fft, Brandt:2021kvt,Bonati:2021vbc,Caristo:2021tbk} and this previous knowledge will help us to fix the parameters of our simulations and will greatly simplify our work.

Since the physical properties of the flux tube are well summarized by its Effective String Theory (EST) description, our ultimate goal 
will be to study the effective string model which describes the reconfined phase and see whether it is akin to the one that has been recently shown to describe very precisely the flux tube in the ordinary confining phase \cite{Caristo:2021tbk}. 
To this end we shall first briefly recall the main features of trace deformed theories and of the EST description of confinement and then present the results of our simulations. A more detailed discussion of our results can be found in \cite{in_preparation}.


\section{The reconfined phase of the $\SU(2)$ gauge theory in (2+1) dimensions.}

We focused our analysis on the $\SU(2)$ gauge theory in (2+1) dimensions for which several results (and in particular the behaviour of the flux tube in the ordinary confining phase \cite{Caristo:2021tbk}) are known with good precision.

We regularize the theory on a finite cubic lattice of spacing $a$ and sizes $aN_t$ in the compactified ``Euclidean-time'' direction and $aN_s$ in the two other (``spatial'') directions. To simplify notations, in the following we will set $a=1$. Periodic boundary conditions are assumed in all directions and we always take $N_s \gg N_t$.

Following~\cite{Bonati:2018rfg,Bonati:2019kmf,Bonati:2020lal,Athenodorou:2020clr} we define the action of the trace deformed model as 
\begin{equation}
S^{\mathrm{def}} = S_{\rm W} + h \sum_{\vec{x}} |P (\vec{x})| ^2\ ,
\label{eq:lattice_action}
\end{equation}
where $S_{\rm W}$ is the usual Wilson action
\cite{Wilson:1974sk}
\begin{equation}
S_{\rm W}= -\frac{\beta}{2} \sum_{x} \sum_{0 \le \mu < \nu \le 2} \Tr U_{\mu\nu} (x)
\end{equation}
and  $P (\vec{x})$ denotes the Polyakov loop.

In the following we shall be mainly interested in the two-point correlation function of Polyakov loops which is defined as
\begin{equation}
\label{def_G}
G(R) = \frac{1}{2 \, {N_s}^2} \; \left\langle \sum_{\substack{\vec{x} \\ k=1, 2}} \ P\left(\vec{x}\right) P^\dagger \left(\vec{x}+R \hat{k}\right) \right\rangle ,
\end{equation}
where the sum is over all spatial coordinates $\vec{x}$ and the two spatial directions $\hat k$.

One of the advantages of studying the $\SU(2)$ model in (2+1) dimensions is that we can leverage previous studies to fix the parameters of the model.  In particular we can use the scale setting expression obtained in \cite{Teper:1998te}   
\begin{equation}\label{scale_setting}
\sqrt{\sigma(\beta)}=\frac{1.324(12)}{\beta}+\frac{1.20(11)}{\beta^2}+\mathcal{O}(\beta^{-3})\ ,
\end{equation}
(which is expected to be valid for $\beta\ge 4.5$) where $\sigma$ denotes the zero-temperature string tension.

The temperature $T$ of the system is given, as usual, by the inverse of the lattice size in the in the $\hat{0}$ direction $T=1/N_t$: as a consequence, $T$ can be varied by changing $N_t$, or the lattice spacing (which can be varied continuously by tuning $\beta$), or both. In the following we shall study the system in the reconfined phase, for several values of $N_t$ and two values of $\beta$ (which will allow us to test the scaling behaviour of our results). To fix precisely the location of the reconfinement transition it will 
 be useful to have precise estimates of the deconfinement temperature $T_c$ for the ordinary $\SU(2)$ gauge theory.
 Very accurate estimates of $T_c$ for various values of $N_t$, can be found in ref.~\cite{Edwards:2009qw}. From these values we extrapolated the critical temperature for the values of $\beta$ used in this paper. Details on our simulations are reported in tab.~\ref{tab1}.

\subsection{The phase diagram}

\begin{figure}[h]
\begin{center}
\includegraphics[width=0.70\textwidth]{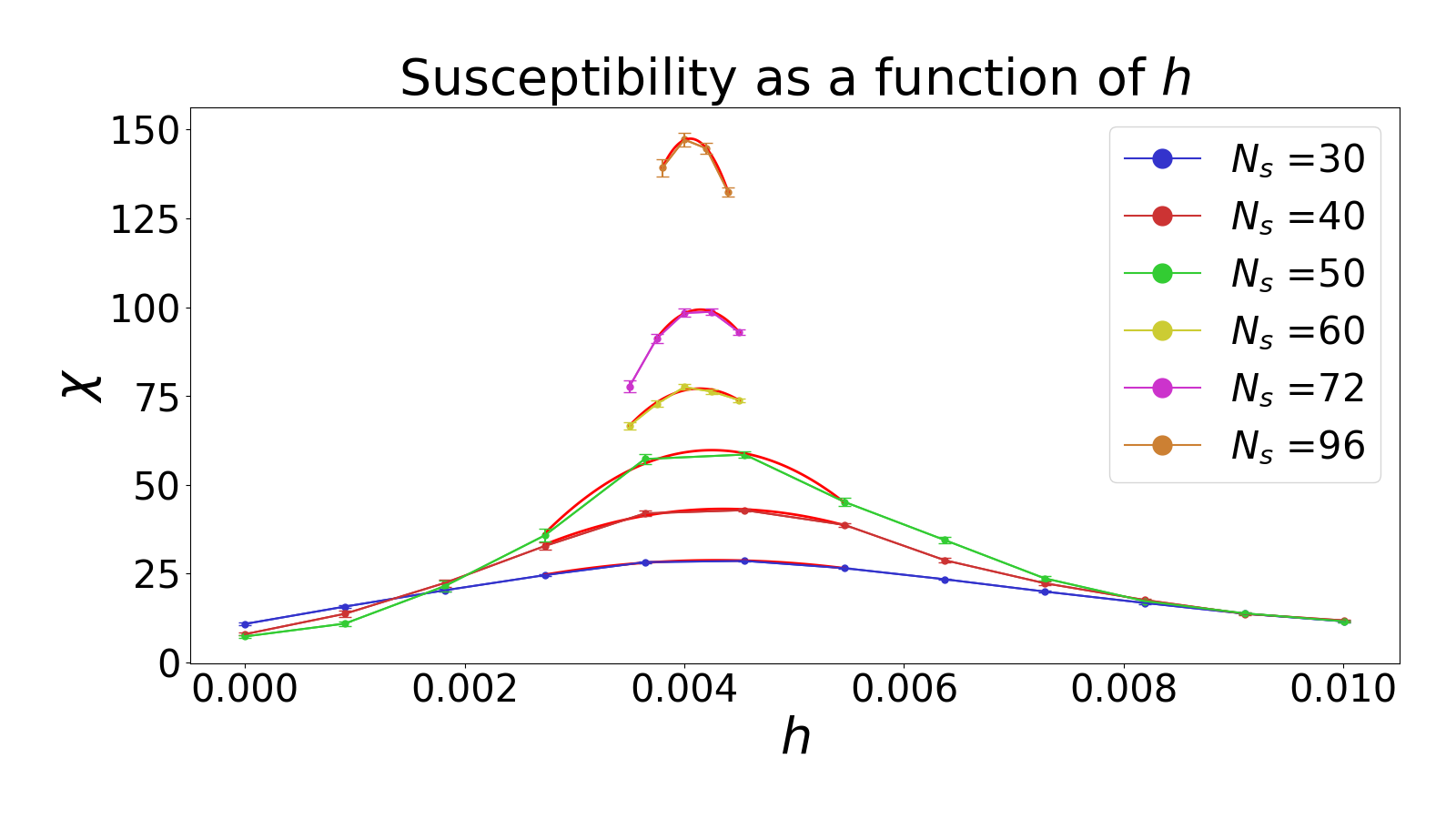}
\caption{Behaviour of the susceptibility for the $\SU(2)$ model in (2+1) dimensions with $N_t=10$ and $\beta=23.3805$, for different values of the lattice extent $N_s$. For this value of $\beta$ the deconfinement transition of the ordinary $\SU(2)$ gauge theory is located at $N_{t,c}=15$, so with $N_t=10$ we are deeply in the deconfined phase.}
    \label{fig1}
\end{center}
\end{figure}

At a fixed value of $\beta$ and $N_t$, chosen so as to have the model in the deconfined phase of the ordinary $\SU(2)$ gauge theory, if we increase $h$ at some point we find a \emph{"reconfinement transition"} which in all the models studied up to now \cite{Bonati:2018rfg,Bonati:2019kmf,Bonati:2020lal,Athenodorou:2020clr}  is always of the first order. To test the nature of the phase transition in our case we performed a finite size scaling analysis of the susceptibility of the Polyakov loop, see fig.~\ref{fig1}. The scaling behaviour of said observable suggests that also in our case we have a first order phase transition. 

\section{The effective string description of the flux tube.}

A general introduction to the subject can be found in
 \cite{Aharony:2013ipa,Brandt:2016xsp,Caselle:2021eir}. We recall here only the main results which will be useful for our analysis.

In the  Effective String Theory (EST) approach the flux tube joining the quark-antiquark pair is described by a thin fluctuating string. In this framework the expectation value of the Polyakov loop correlator is proportional to the free energy of this string. This description is valid only above a critical distance $R_c$ between the quark and the antiquark and for this reason it only represent an effective "low energy" description of confinement, however in its range of validity it describes impressively well the correlator, with an almost perfect agreement between EST predictions and LGT data (see the reviews \cite{Aharony:2013ipa,Brandt:2016xsp,Caselle:2021eir}). 

The EST description becomes simpler when $R \gg N_t$ i.e. in the high temperature limit (just below the deconfinement transition, but still in the confining phase). In this limit the so called "boundary terms" (associated to the self-energy of the quarks) can be neglected \cite{Caselle:2021eir} and the EST prediction assumes a particularly simple form \cite{Luscher:2004ib}, which is valid for any possible EST and in (2+1) dimension is: 
        \begin{equation}
            G(R) = \sum_{n=0}^\infty |v_n(N_t)|^2\frac{E_n}{\pi}K_{0}(E_nR)
        \end{equation}
where $K_0$ is modified Bessel function of order zero,  $E_n$ is the energy of the $n$-th excited state of the string and $v_n(N_t)$ its amplitude (which also encodes the multiplicity of the $n$-th energy level).
In the large $R$ limit the sum is dominated by the lowest state and 
thus we may approximate the Polyakov loop correlators as:
\begin{equation}
    G(R) = A(N_t) \; K_{0} \big( E_0(N_t) \, R \big)
    \label{E0}
\end{equation}
from which it is possible to extract the ground state energy $E_0(N_t)$ and $A(N_t)$ via a fitting procedure.

Equation \eqref{E0} holds for any EST. If we want to characterize the particular EST which describes our LGT we must look at the $N_t$ dependence of $E_0(N_t)$ and of the amplitude $A(N_t)$, which allow to distinguish between different effective strings.
        
For instance, for the {Nambu-Goto (NG) string} we have the well known result:
\begin{equation}
    E_0=\sigma N_t \sqrt{1-\frac{\pi}{3\sigma N_t^2}}
    \label{NG}
\end{equation}

In the following we shall use this strategy to identify the EST describing the reconfined phase of the model.

\subsection{EST of the ordinary confining phase of the $\SU(2)$ LGT in (2+1) dimensions.}
In the case of the ordinary confining phase of the model the Nambu-Goto expression of eq.~(\ref{NG}) gives a very good approximation of the actual EST.  This good agreement is due to the so called low-energy universality \cite{Aharony:2013ipa,Dubovsky:2012sh} and has been recently confirmed in \cite{Caristo:2021tbk} and \cite{Caselle:2024zoh} where the tiny deviations with respect to the Nambu-Goto prediction were detected and evaluated using high precision Monte Carlo simulations. This agreement can be appreciated looking at fig.~\ref{fig2} where the blue open circles are the results of simulations in the ordinary $\SU(2)$ model (taken from ref.~\cite{Caristo:2021tbk}) and the continuous line is the Nambu-Goto prediction of eq.~(\ref{NG}): Only the points nearest to the deconfinement transition show a deviation from the Nambu-Goto prediction.

\section{Results}

\subsection{Analysis of the Polyakov loop correlators in the reconfined phase.}

To analyse the behaviour of the flux tube in the reconfined phase we selected a set of values of $\beta,N_t$ and $h$ beyond the reconfinement transition.

\begin{table}[ht]
\centering %
\begin{tabular}{| c | c | c | c | c | c |} 
\hline
$\beta$ & $N_{t,c}$ & h & $N_t$ & $N_s$ \\ \hline
 $23.3805$ & 15 & 0.005 &  9,10,11,12,13,14           & 96 \\ \hline
 $23.3805$ & 15 & 0.006 &  8,9,10,11,12,13,14         & 96 \\ \hline
 $23.3805$ & 15 & 0.007 &  7,8,9,10,11,12,13,14       & 96 \\ \hline
$27.4745$  & 20 & 0.004 &  11,12,13,14,15,16,17,18,19 & 96 \\ \hline
$27.4745$  & 20 & 0.005 &  9,10,11,12,13,14,15,16,17  & 96 \\ \hline

\end{tabular}
\caption{Some information on the simulations.}
\label{tab1}
\end{table}
We simulated the model for the values of $\beta,h$ and $N_t$ listed in tab.~\ref{tab1}.
For each of these values we measured the correlator $G(R)$ for all the values $3\leq R \leq 23$.
We fitted the $R$ dependence of the correlator with the effective string expectation of eq.~(\ref{E0}), modified so as to keep into account the periodic boundary conditions:  
\begin{equation}\label{eq:fS}
	G(R)=A \Big[ K_0 \big(E_0 \, R \big) + K_0 \big(E_0 \, (N_s - R) \big) \Big] \, .
\end{equation} 
As mentioned above this expression is expected to describe the large distance behaviour of the flux tube without any specific assumption on the effective string model. In our case we always found a good $\chi^2$ when fitting the data in the range $15\leq R \leq 23$. An example of our results, for $\beta=23.3805$ and $h=0.005$ is reported in tab.~\ref{tab:E0Nt005}.


It is easy to see from the fig.~\ref{fig2} that the data (green open squares) completely disagree with the EST picture discussed above, which is instead valid in the ordinary confining phase (blue open circles).
This shows that the two confining mechanisms must be different and prompts us to try to find a different EST, able to describe the model in its reconfined phase. An interesting candidate to play this role is the so called "rigid" string which we shall discuss in the next section.

\subsection{The "rigid" string}

The {"rigid string"} is obtained
adding to the Nambu-Goto action a term proportional to (the square of) the extrinsic curvature which has the effect of increasing the {"stiffness"} of the string.
For a world-sheet $X^\mu (\xi_0, \xi_1)$, on which is induced the metric $g_{ab} = \partial_a X^\mu \partial_b X_\mu$, the rigid string action reads
\begin{equation}
S_{R}= \int_\Sigma d^2\xi \, \sqrt{g} \, \left[\sigma  + \gamma_2 {\cal K}^2 +  \dots \right] \, ,
\label{rigid}  
\end{equation}
where {${\cal K}$} is the extrinsic curvature defined as ${\cal K}=\Delta (g) X$, with  
\begin{equation}
\Delta(g)=\frac{1}{\sqrt{g}} \, \partial_a \, \left[\sqrt{g} \, g^{ab}\partial_b \right] \, .
\end{equation}

In the "physical gauge" in (2+1) dimensions, keeping only the first order terms this boils down to:
{
\eq
S_{R}= \int_\Sigma d^2\xi \; \left[ \sigma \, \partial X \partial X + \gamma_2 \, \partial^2 X \partial^2 X + \dots \right]  \, ,  
\en
}

The action in eq.~(\ref{rigid}) has a long history. Originally introduced to describe the physics of fluid membranes~\cite{Peliti:1985eo,Helfrich:1985eo,Forster:1986ot}, it was later proposed by Polyakov and by Kleinert as a way to stabilize the Nambu-Goto action~\cite{Polyakov:1986cs, Kleinert:1986bk}.

The standard approach to study the model was to treat the {${\cal K}^2$} term as a perturbation of the gaussian (Nambu-Goto) one (see for instance~\cite{Braaten:1986bz} and~\cite{German:1989vk}). However the interquark potential computed in simulations of non Abelian LGTs, in both (2+1) and (3+1) dimensions, turned out to be in substantial agreement with the prediction of an EST whose action only contains the NG term. This suggests an almost negligible value of the rigidity correction.
In recent years it has been understood that this is due to the so called "low energy universality"  \cite{Aharony:2013ipa,Dubovsky:2012sh}  i.e. to the fact that the  {${\cal K}^2$} term can be eliminated since it is proportional to the eq. of motion of the NG string and that a  better proposal to describe higher order perturbations to the Nambu-Goto action is instead
\begin{equation}
S_{BNG}= \int_\Sigma d^2\xi \, \sqrt{g} \,\left[\sigma +  \gamma_3 {\cal K}^4 + \dots \right] \, .
\nonumber
\end{equation}

This term is responsible for the tiny deviations with respect to the Nambu-Goto predictions in ordinary confining gauge theories that we mentioned above and which were recently observed in high precision simulations of (2+1) dimensional  $\SU(N)$ LGTs \cite{Caristo:2021tbk,Caselle:2024zoh} and are represented in fig.~\ref{fig2} by the blue dashed line.
However this is not the end of the story.

\subsection{The Polchinski-Yang solution}

A completely different approach to study $S_R$ was proposed in 1992 by Polchinski and Yang which described the rigid string {assuming the quartic term as the dominant one and the quadratic NG term as a small perturbation}. 
This corresponds to a completely different vacuum and requires {$\gamma_2 \gg N_t^2\sigma$} and {$N_t^2\sigma \ll 1$}, resulting (apparently) in an unphysical regime since for {\bf ordinary YM theories} {$N_{t,c}\sim 1/\sqrt{\sigma}$}. Consequently, if {$N_t^2\sigma \ll 1$}, the model is deconfined and thus we do not expect the presence of a confining flux tube to justify an effective string description.
Despite this, this regime was studied in great detail for completely different reasons.  The goal was to show that in this particular regime the  (unphysical) high temperature behaviour of model was the same of that of QCD in the large $N$ limit.

In the following we shall be interested in the second regime.  Our claim is that even if it is  unphysical in ordinary YM models it is instead realized in the trace deformed YM models for large enough values of $h$. The phase transition which is observed in the phase diagram of these model is exactly the phase transition which separates the two regimes described above. 
We are interested in particular to the behaviour of the ground state energy of the string $E_0$ defined as $E_0=-(\log Z)/R$ where $Z$ is the partition Function of the string. Following \cite{Polchinski:1992ty} we have (for generic values of the transverse dimensions):
\begin{equation}
E_0 = w \, \lambda \, ,
\end{equation}
where
\begin{equation}
    w=\sqrt{N_t^2-\frac{(d-2)N_t}{2}\sqrt{\frac{1}{2\gamma_2\lambda}}}
\end{equation}
and
\begin{equation}
\sqrt{\lambda}=\frac{3}{8}\frac{(d-2)}{N_t \sqrt{2\gamma_2}}+\sqrt{\frac{9}{128}\frac{(d-2)^2}{\gamma_2 N_t^2}+{\sigma}-\frac{\pi(d-2)}{3N_t^2}}
\end{equation}
(see also \cite{Ambjorn:2014rwa} for a slightly different formulation).

\begin{figure}
\begin{center}
\includegraphics[width=0.70\textwidth]{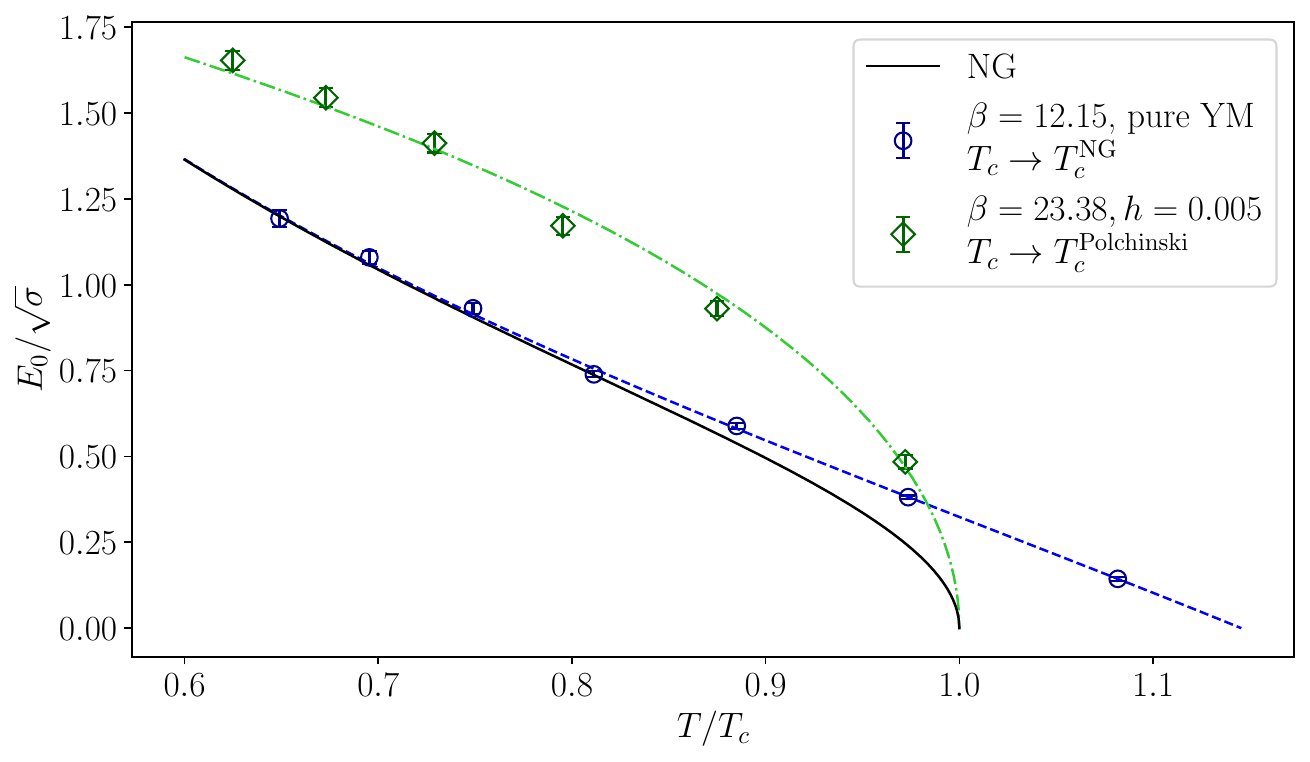}
\caption{Values of the adimensional ratio $E_0/\sqrt{\sigma}$ as a function of $T/T_c$. The blue open circles are taken from \cite{Caristo:2021tbk} and correspond to the ordinary confining $\SU(2)$ model, while the green open squares are the data discussed in the present paper for the reconfined phase. The continuous line is the Nambu-Goto prediction, the blue dashed line describes the Nambu-Goto prediction plus the term proportional to ${\cal K}^4$ discussed in \cite{Caristo:2021tbk} while the green dash-dotted curve is the Polchinski-Yang solution.}
\label{fig2}
\end{center}
\end{figure}

\begin{figure}
\begin{center}
\includegraphics[width=0.70\textwidth]{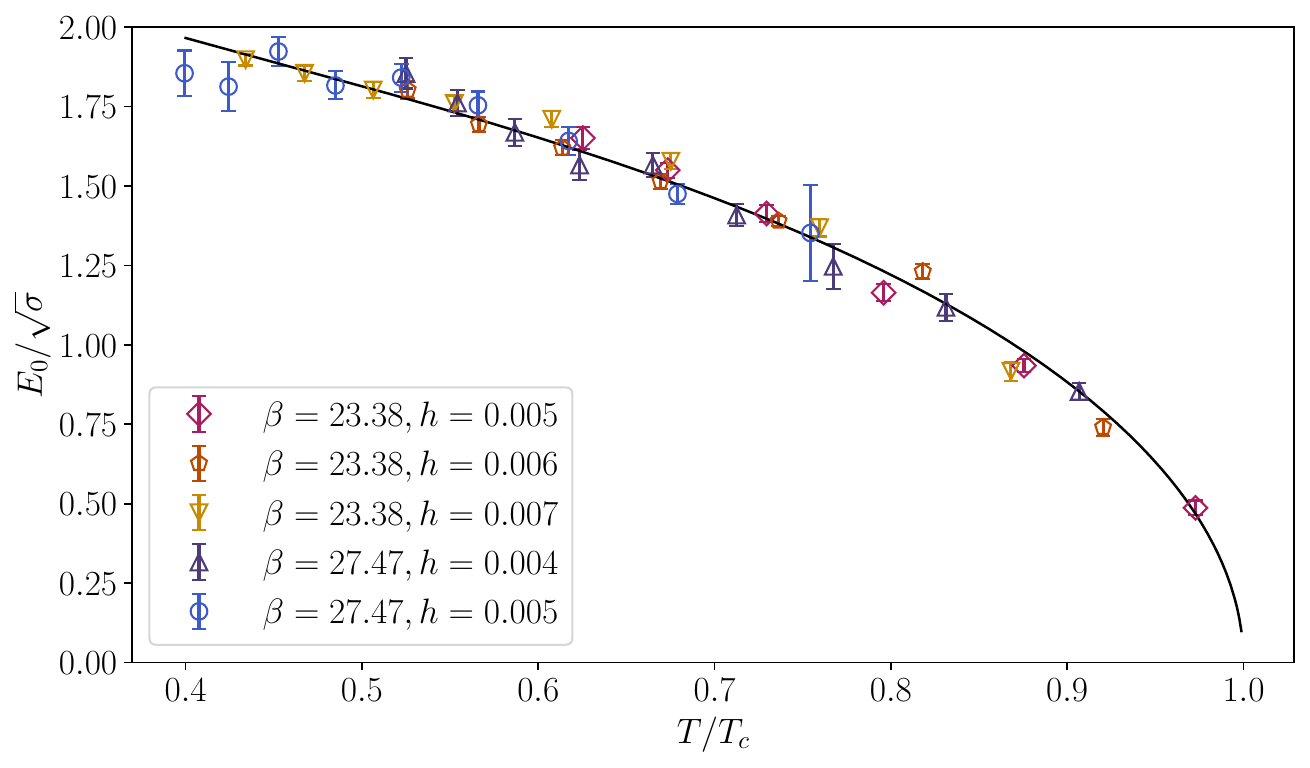}
\caption{Data for all  values of $\beta$ and $h$ that we studied, compared with the Polchinski-Yang solution.}
\label{fig4}
\end{center}
\end{figure}

This expression depends on only two degrees of freedom: $\sigma$ and $\gamma_2$.  Fitting these two parameters with our data we find the is the green dash-dotted line of fig.~\ref{fig2}, which agrees remarkably well with the results of the simulations.
This agreement is confirmed also for the other values of $h$ and $\beta$ that we studied (see fig.~\ref{fig4}). 
Further details on this analysis and on other properties of the reconfined flux tube like its width and shape will be discussed in a forthcoming paper \cite{in_preparation}. 

\begin{table}[h]
	\centering
	\begin{tabular}{|c|c|c|}
		\hline
		$(N_t\times N_s^2,h)$  & $A$       & $E_0$     \\ \hline
		$( 9\times96^2,0.005)$ & 0.0330(5) & 0.0125(5) \\ \hline
		$(10\times96^2,0.005)$ & 0.0349(4) & 0.0240(6) \\ \hline
		$(11\times96^2,0.005)$ & 0.0357(5) & 0.0302(7) \\ \hline
		$(12\times96^2,0.005)$ & 0.0367(5) & 0.0364(7) \\ \hline
		$(13\times96^2,0.005)$ & 0.0366(5) & 0.0398(7) \\ \hline
		$(14\times96^2,0.005)$ & 0.0362(5) & 0.0426(7) \\ \hline
	\end{tabular}
	\caption{Results from the fit of $G(R)$ with eq.~\eqref{eq:fS} for $\beta = 23.3805$ and $h = 0.005$.}
	\label{tab:E0Nt005}
\end{table}
  
\subsection*{Acknowledgements}

This work has been supported by the Italian PRIN ``Progetti di Ricerca di Rilevante Interesse Nazionale – Bando 2022'' prot. 2022TJFCYB, and by the ``Simons Collaboration on Confinement and QCD Strings'' funded by the Simons Foundation. The simulations were run on CINECA computers. C.~Bonati acknowledges support from the NPQCD Scientific Initiative of the Italian Nuclear Physics Institute (INFN), while M.~Caselle, D.~Panfalone and L.~Verzichelli acknowledge support from the SFT Scientific Initiative of the same institution.

\bibliographystyle{JHEP}
\providecommand{\href}[2]{#2}\begingroup\raggedright\endgroup

\end{document}